\documentclass[twocolumn, journal, a4paper, final]{IEEEtran}
\IEEEoverridecommandlockouts

\usepackage{color}
\usepackage{bm}
\usepackage{graphicx}
\usepackage{amsmath}
\usepackage{amssymb}
\usepackage{algorithm}
\usepackage{algorithmic}
\usepackage{multirow}
\usepackage{booktabs}
\usepackage{array}
\usepackage{lipsum}
\usepackage{enumerate}
\usepackage{stfloats}
\usepackage{subfigure}
\usepackage{cases}
\usepackage{diagbox}
\usepackage{cite}
\usepackage{enumitem}
\usepackage{balance}
\usepackage{geometry}
\newtheorem{remark}{Remark}

\newtheorem{lemma}{Lemma}

\allowdisplaybreaks[4]

\title{ Enhancing Physical Layer Security in Dual-Function Radar-Communication Systems with Hybrid Beamforming Architecture}
\author{
	Lingyun Xu, \IEEEmembership{Graduate Student Member~IEEE,}
	Bowen Wang, \IEEEmembership{Graduate Student Member~IEEE,} \\
	Huiyong Li  and Ziyang Cheng,~\IEEEmembership{Member~IEEE}
	\vspace{-1.5em}
	
	\thanks{Note: This paper has been published by IEEE WCL and is the \textbf{full-length} paper. We will keep this paper updated if we find any mistakes.}
	\thanks{
		Manuscript received 24 November 2023; revised 19 March 2024; accepted 21 March 2024. AQ1 This work was supported in part by the National Natural Science Foundation of China under Grant 62001084, Grant 62231006, and Grant 62371096; in part by the Sichuan Science and Technology Program under Grant 2023NSFSC1385; and in part by the Municipal Government of Quzhou under Grant 2022D013. The associate editor coordinating the review of this article and approving it for publication was D. Mishra.
		(\emph{Corresponding author: Ziyang Cheng})
	}
	\thanks{Lingyun Xu, Bowen Wang, and Ziyang Cheng are with the School of Information and Communication Engineering, University of Electronic Science and Technology of China, Chengdu 610054, China (e-mail: xusherly@163.com; bwwang@std.uestc.edu.cn; zycheng@uestc.edu.cn).
		
		Huiyong Li is with the School of Information and Communication Engineering, University of Electronic Science and Technology of China, Chengdu 610054, China, and also with the Yangtze Delta Region Institute, University of Electronic Science and Technology of China, Quzhou 324003, China (e-mail: hyli@uestc.edu.cn).
		
		Digital Object Identifier 10.1109/LWC.2024.3382035}
}

\begin{document}
	
	\maketitle
	
	\begin{abstract}
		In this letter, we investigate enhancing the physical layer security (PLS) for the dual-function radar-communication (DFRC) system with hybrid beamforming (HBF) architecture, where the base station (BS) achieves downlink communication and radar target detection simultaneously.
		{We consider an eavesdropper intercepting the information transmitted from the BS to the downlink communication users with imperfectly known channel state information.}
		{Additionally, the location of the radar target is also imperfectly known by the BS.}
		To enhance PLS in the considered DFRC system, we propose a novel HBF architecture, which introduces a new integrated sensing and security (I2S) symbol.
		The secure HBF design problem for DFRC is formulated by maximizing the minimum legitimate user communication rate subject to radar signal-to-interference-plus-noise ratio, eavesdropping rate, hardware and power constraints.
		To solve this non-convex problem, we propose an alternating optimization based method to jointly optimize transmit and receive beamformers.
		Numerical simulation results validate the effectiveness of the proposed algorithm and show the superiority of the proposed I2S-aided HBF architecture for achieving DFRC and enhancing PLS.
	\end{abstract}
	
	
	\section{Introduction}
	
	With the exponentially increasing number of wireless communication devices, spectrum resources become scarce \cite{zhang2021overview}.
	To improve spectrum efficiency, spectrum sharing between communication and radar has drawn much attention \cite{Hass2019DFRC}.
	This motivates research interests in dual-function radar-communication (DFRC) systems \cite{Liu2022DFRC}, which can achieve radar sensing and wireless communication simultaneously, adopting the same signaling scheme, operating on the same platform, and sharing the same spectrum \cite{Liu2020HBFDFRC}.
	
	Although DFRC systems have many benefits, the broadcasting nature of wireless media and the spectrum-sharing mechanism make DFRC vulnerable to security threats \cite{Shi2023survey}.
	Typically, the DFRC waveform is designed to concentrate power in the direction of radar targets to enhance sensing performance \cite{wei2022toward}, causing communication information contained in the probing signal vulnerably eavesdropped by the target.
	Facing these security challenges, some investigations have been conducted to enhance physical layer security (PLS) for DFRC systems \cite{Deli2018SeDFRC, Chu2021SeDFRC, Su2021SeDFRC, Su2022SeDFRC, Dong2023SeDFRC}.
	For instance, authors in \cite{Deli2018SeDFRC} investigate waveform design for DFRC system with single communication user and single target as potential eavesdropper, where the eavesdropping rate is combated by simultaneously transmitting desired information and false information to confuse the eavesdropper.
	As a further step, an artificial noise (AN) aided secure beamforming design method is proposed in \cite{Chu2021SeDFRC} to disrupt eavesdropper reception in a multi-user and single-target scenario, where the eavesdropping signal-to-interference-plus-noise ratio (SINR) is minimized while guaranteeing SINR of legitimate users and transmit beampattern similarity.
	Furthermore, a robust waveform design for DFRC system with imperfect channel state information (CSI) is considered in \cite{Su2021SeDFRC}, where the eavesdropping SINR is minimized while guaranteeing SINR at legitimate users.
	Besides, authors in \cite{Dong2023SeDFRC} jointly design DFRC waveform in the multi-user and multi-target scenario, where the radar signals are utilized to jam the eavesdroppers.
	
	It is noteworthy that most existing works employ the fully-digital architecture to realize PLS for DFRC systems.
	Although these schemes can improve the performance of PLS to some extent, they only work efficiently in sub-6G Hz and lead to tremendous power consumption in DFRC systems with massive MIMO.
	To support millimeter wave (mmWave) DFRC with massive MIMO, hybrid beamforming (HBF) design for DFRC has been studied in \cite{Liu2020HBFDFRC, cheng2022hybrid, cheng2022double}, which utilizes phase shifters to construct the analog beamformer (ABF) and fewer radio frequency (RF) chains to link the digital beamformer (DBF) to save hardware costs and power consumption \cite{heath2016overview}.
	However, the mentioned works \cite{Liu2020HBFDFRC, cheng2022hybrid, cheng2022double} do not account for PLS requirements in the DFRC system design. 
	Furthermore, the conventional HBF architecture is tailored for mmWave communication, and its suitability for achieving effective PLS remains unexplored.
	Therefore, it is crucial to investigate boosting PLS in mmWave DFRC with HBF architecture from architecture to design mechanism perspectives.
	Besides, differentiating from most existing works that regard the radar target as an eavesdropper in ideal scenarios with perfect CSI, this letter explores a more comprehensive and practical scenario, where the BS only knows imperfect eavesdropper CSI and radar target location.
	
	To fill in this gap, we develop a novel HBF architecture to enhance PLS for DFRC.
	Specifically, we introduce an integrated sensing and security (I2S) symbol in DBF, which not only jams the eavesdropper but also guarantees radar detection performance.
	Then, we formulate the secure I2S-aided HBF design problem to maximize the lowest legitimate communication user data rate, while suppressing eavesdropping rate, guaranteeing radar SINR, power and hardware constraints.
	To solve this non-convex problem, an alternating optimization (AltOpt) based algorithm is proposed.
	Finally, we demonstrate numerical simulation results to validate the effectiveness of the proposed secure HBF algorithm and demonstrate the superiority of the proposed I2S-aided HBF structure over conventional HBF structure in terms of enhancing PLS.
	
	\textit{Notation:} 
	${\bf{A}}(i,j)$ denotes the $(i,j)$-th element of the matrix $\bf A$. 
	$(\cdot)^T$ and $(\cdot)^H$ denote transpose and conjugate transpose operators, respectively. 
	${\|  \cdot  \|_F}$ and ${\rm Tr}({\cdot})$ denote Frobenius norm and trace, respectively.
	${\mathbb E}\{   \cdot \}$ and $\Re \{  \cdot  \}$ denote expectation and the real part of a complex-valued number, respectively. $[\cdot ]^{+}$ denotes ${\rm max} \{ \cdot,0\}$.

	\section{System Model and Problem Formulation}
	
	\begin{figure}[t]
		\centering
		\subfigure[]{
			\includegraphics[width=0.488 \linewidth]{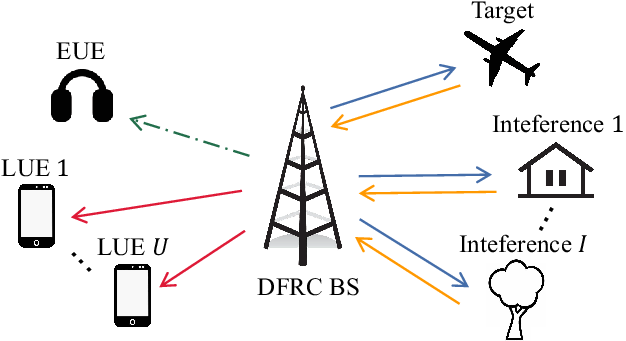} 
			\label{scene}
		}
		\hspace{-1.6em}
		\subfigure[]{
			\includegraphics[width=0.488 \linewidth]{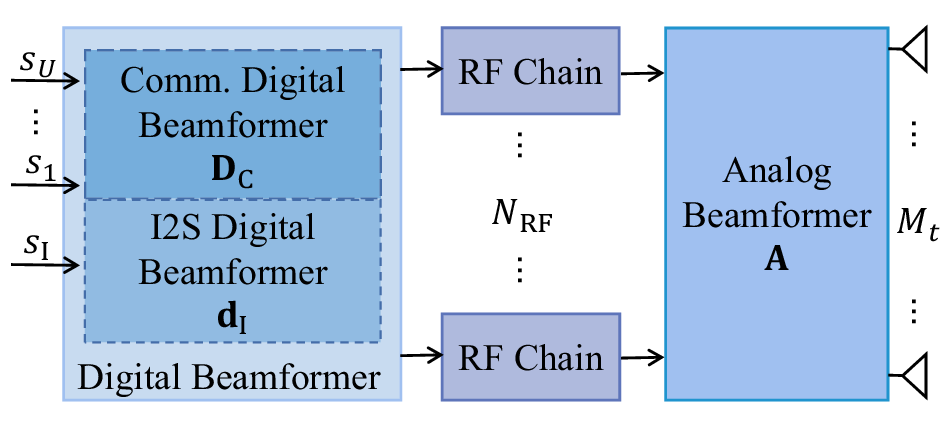} 
			\label{DFRC_BS}
		}
		\caption{(a) A DFRC system with a potential eavesdropper. (b) The proposed DFRC BS with I2S-aided HBF at the transmitter.}
		\label{sys_model}
	\end{figure}
	
	As shown in Fig.~\ref{scene}, we consider a DFRC system equipped with HBF architecture that serves $U$ single-antenna legitimate communication users, noted as LUE, and detects a radar target in the presence of $I$ signal-dependent interference sources.
	There is also a single-antenna potential eavesdropper, noted as EUE, which attempts to intercept confidential data sent from DFRC BS to LUEs.
	The radar receiver co-located with the transmitter collects echo to detect the target.
	
	\subsection{Transmit Architecture}
	
	In this work, to achieve secure DFRC, we propose a novel I2S-aided HBF architecture as shown in Fig.~\ref{DFRC_BS}, composing data stream, DBF, and ABF. 
	To clarify, the operating mechanism of the proposed I2S-aided HBF is outlined as follows:
	
	\textit{Stage 1:} The digital baseband generates two data streams, i.e., ${{\bf{s}}_{\rm{C}}}\! =\! {[ {{s_1}, \ldots ,\;{s_U}} ]^T}\! \in \! {\mathbb{C}^{U}}$ and $s_{\rm{I}}$. 
	Specifically, ${{\bf{s}}_{\rm{C}}}$ denotes communication data symbols, assuming $\mathbb{E} \{ {{\bf{s}}_{\rm{C}}}{\bf{s}}_{\rm{C}}^H \} = {\mathbf{I}}_U $. 
	$s_{\rm{I}}$ denotes the I2S symbol, which is assumed ${s_{\rm{I}}}\sim \mathcal{C}\mathcal{N}(0,1)$.
	
	\textit{Stage 2:} Two data streams ${{\bf{s}}_{\rm{C}}}$ and $s_{\rm{I}}$ are separately  processed by the communication DBF ${\bf{D}}_{\rm{C}} = \left[ {{{\bf{d}}_{{\rm{C,}}1}}, \ldots ,\;{{\bf{d}}_{{\rm{C,}}U}}} \right] \in {\mathbb{C}^{{N_{{\rm{RF}}}} \times U}}$ and the I2S DBF ${{\bf{d}}_{{\rm{I}}}} \in {\mathbb{C}^{{N_{{\rm{RF}}}}}}$, respectively.
	Then, the input signal at RF chains after digital-to-analog conversion is given by ${\bf{x}}_{\rm{A}} = {{\bf{D}}_{{\rm{C}}}}{{\bf{s}}_{\rm{C}}} + {{\bf{d}}_{{\rm{I}}}}{s_{\rm{I}}}$.
	
	\textit{Stage 3:} The analog signal is up-converted by RF chains. 
	Following that, the high-frequency analog signal is modulated by the ABF ${{\bf{A}}} \in {\mathbb{C}^{{M_t} \times {N_{\rm RF}}}}$, which is implemented by phase shifters.
	Without loss generality, we assume the ABF $\bf{A}$ adopts the fully-connected architecture, i.e., $\left| {{\bf{A}}[i,j]} \right| = 1,\forall i,j$.
	
	According to above operating mechanism, the transmitted signal ${\bf{x}}_{\rm{TX}} \in {\mathbb{C}^{{M_t}}}$ at the DFRC BS is given by
	\begin{equation}
	{\bf{x}}_{\rm{TX}} = {\bf{A}}{\bf{x}}_{\rm{A}} = {\bf{A}}({{\bf{D}}_{{\rm{C}}}}{\bf{s}}_{\rm{C}} + {{\bf{d}}_{{\rm{I}}}}{s_{\rm{I}}})
	\label{1}
	\end{equation}
	
	\begin{remark}
		Unlike conventional HBF architecture, the proposed I2S-aided HBF architecture introduces the I2S signal ${s_{\rm{I}}}$.
		The benefits of introducing I2S signal ${s_{\rm{I}}}$ can be summarized as:
		1) From the secure communication side, I2S signal ${s_{\rm{I}}}$ can be viewed as AN, which plays a significant role in disturbing eavesdropper and ensuring security.
		2) From the radar side, I2S signal ${s_{\rm{I}}}$ can be viewed as a dedicated radar waveform, which improves the degree of freedom of radar waveform design.
	\end{remark}

	\subsection{Communication Model}
	
	After propagating through the mmWave channel, the received signals of the $u$-th LUE and EUE are expressed as
	\begin{align}
	&{y_u} = {\bf{h}}_u^H{\bf{A}}{{\bf{d}}_{{\rm{C,}}u}}{s_u}  + \!\!\!\!\textstyle \sum\limits_{i = 1,i \ne u}^U \!\!\!\! {{\bf{h}}_u^H{{\bf{A}}}{{\bf{d}}_{{\rm{C,}}i}}{s_i}}  +  {\bf{h}}_u^H{{\bf{A}}{{\bf{d}}_{{\rm{I}}}}}{s_{\rm{I}}}  +  {n_u} 
	\label{2}\\
	&{y_{{\rm{e}}}}  =  {\bf{h}}_{{\rm{e}}}^H{{\bf{A}}}{{\bf{d}}_{{\rm{C,}}u}}{s_u}  + \!\!\!\! \textstyle\sum\limits_{i = 1,i \ne u}^U \!\!\!\! {{\bf{h}}_{{\rm{e}}}^H{{\bf{A}}}{{\bf{d}}_{{\rm{C,}}i}}{s_i}} + {\bf{h}}_{{\rm{e}}}^H{{\bf{A}}{{\bf{d}}_{{\rm{I}}}}}{s_{\rm{I}}}  + {n_{\rm{e}}} \label{3}
	\end{align}
	where ${\bf{h}}_u$ and ${\bf{h}}_{{\rm{e}}}$ denotes the mmWave channels for the $u$-th LUE and EUE, respectively.
	${n_u} \sim {\cal C}{\cal N}( {0,\;\sigma _{u}^2} )$ and ${n_{\rm{e}}} \sim {\cal C}{\cal N}( {0,\;\sigma _{\rm{e}}^2})$ represent the additive white Gaussian noise (AWGN) of the $u$-th LUE and EUE, respectively.
	
	Based on \eqref{2}, the SINR at the $u$-th LUE is then given by
	\begin{equation}
	{\rm{SIN}}{{\rm{R}}_u} = \frac{{{{\left| {{\bf{h}}_u^H{{\bf{A}}}{{\bf{d}}_{{\rm{C,}}u}}} \right|}^2}}}{{\sum\nolimits_{i = 1,i \ne u}^U {{{\left| {{\bf{h}}_u^H{{\bf{A}}}{{\bf{d}}_{{\rm{C,}}i}}} \right|}^2}}  + {{\left| {{\bf{h}}_u^H{{\bf{A}}}{{\bf{d}}_{{\rm{I}}}}} \right|}^2} + \sigma _u^2}}
	\end{equation}
	
	In this letter, we assume that EUE has infinite computation capability to eliminate multi-user interference. 
	Thus, based on \eqref{3}, the SINR at EUE about the $u$-th LUE is given by
	\begin{equation}
	{\rm{SINR}}{_{{\rm{e}},u}} = \frac{{{{| {{\bf{h}}_{{\rm{e}}}^H{{\bf{A}}}{{\bf{d}}_{{\rm{C,}}u}}} |}^2}}}{{{{| {{\bf{h}}_{{\rm{e}}}^H{{\bf{A}}}{{\bf{d}}_{{\rm{I}}}}} |}^2} + \sigma _{\rm{e}}^2}}
	\end{equation}
	
	Then, the achievable rates of the $u$-th LUE and EUE can be calculated as
	\begin{equation}
	\!\!\!{{\rm{Rate}}_u}\! = \!{\log _2}\left( {1 \!+\! {\rm{SIN}}{{\rm{R}}_u}} \right) 
	,\;{{\rm{{Rate}}}_{{\rm{e}},u}} \!= \!{\log _2}\left( {1 \!+\! {\rm{{SINR}}}{_{{\rm{e}},u}}} \right) \
	\end{equation}
	Accordingly, the achievable secrecy rate (SR) is defined as
	\begin{equation}
	{\rm{SR}} = \textstyle \sum\nolimits_{u = 1}^U {\rm{SR}}_u= \textstyle \sum\nolimits_{u = 1}^U {{{\left[ {{\rm{Rat}}{{\rm{e}}_u} - {{{{\rm{Rate}}} }_{{\rm{e}},u}}} \right]}^ + }} 
	\label{eq:SR}
	\end{equation}

	\subsection{Radar Model}
	
	Suppose there is a target and $I$ stationary signal-dependent interference sources, i.e., buildings and trees, where the target and the $i$-th interference are located at $\theta _0$ and $\theta _i$, respectively.
	Given the transmit signal ${\bf{x}}_{\rm{TX}}$, the received signal of the radar receiver is given by
	\begin{equation}
	{{\bf{r}}} = {\varsigma _0}{{\bf{a}}_r}({\theta_0}){\bf{a}}_t^H({\theta_0}){\bf{x}}_{\rm{TX}} + \textstyle\sum\nolimits_{i = 1}^I {{\varsigma _i}{{\bf{a}}_r}({\theta _i}){\bf{a}}_t^H({\theta _i}){\bf{x}}_{\rm{TX}}} + {\bf{z}}
	\label{8}
	\end{equation}%
	where ${\bf{z}} \! \sim \! {\cal C}{\cal N}(0,{\sigma^2_r}{\bf I}_{M_r}) \! \in \! {\mathbb{C}^{{M_r}}}$ denotes the AWGN, ${\varsigma _0}$ and ${\varsigma _i}$ are the complex amplitudes of target and the $i$-th clutters, ${\varsigma _i} \!\!  \sim \!\!  \mathcal{CN}( {0,\xi _i^2} )$.
	${{\mathbf{a}}_t}(\theta ) \!\! =  \!\! {[1,{e^{\jmath 2\pi d\sin \theta /\lambda }}, \ldots ,{e^{\jmath 2\pi ({M_t} - 1)d\sin \theta /\lambda }}]^T} \in {\mathbb{C}^{{M_t}}}$ and ${{\mathbf{a}}_r}(\theta ) \!\!  = \!\! {[1,{e^{\jmath 2\pi d\sin \theta /\lambda }}, \ldots ,{e^{\jmath 2\pi ({M_r} - 1)d\sin \theta /\lambda }}]^T} \in {\mathbb{C}^{{M_r}}}$ are the transmit and receive steering vectors in uniform linear arrays (ULAs) at BS, respectively.

	The received signal \eqref{8} at the BS is processed by a receive filter ${\bf{w}} \in {\mathbb{C}^{{M_r}}}$, and the output radar SINR\footnote{For the radar hypothesis test, with generalized likelihood ratio test (GLRT) detector,
		the detection probability $P_d$ of the target can be expressed as ${P_d} = Q( {\sqrt {2{\rm{SINR_r}}} ,\sqrt { - 2\ln ({P_{fa}})} } )$ \cite{De2009radar}, where $Q( \cdot , \cdot )$ is the Marcum Q function of order 1 and ${P_{fa}}$ is the false alarm probability. Thus, the maximization of $P_d$ is equivalent to the maximization of radar SINR, with a specified $P_{fa}$.} can be given by
	\begin{equation}
	{\rm{SIN}}{{\rm{R}}_{{\rm{r}}}} = \frac{{\| {{\varsigma _0}{{\bf{w}}^H}{\bf{\tilde A}}({\theta _0}){{\bf{A}}}{{\bf{D}}_{\rm{CI}}}} \|_F^2}}{{\sum\nolimits_{i = 1}^I {\| {{\varsigma _i}{{\bf{w}}^H}{\bf{\tilde A}}({\theta _i}){{\bf{A}}}{\bf{D}}_{\rm{CI}}} \|_F^2}  + {{| {{{\bf{w}}^H}{\bf{z}}} |}^2}}}
	\label{9}
	\end{equation}
	where ${\bf{\tilde A}}({\theta}) = {{\bf{a}}_r}({\theta}){\bf{a}}_t^H({\theta})$ represents the radar channel, and defining ${\bf{D}}_{\rm{CI}} = [{\bf{d}}_{\rm{I}}, {\bf{D}}_{\rm{C}}]$.

	\subsection{Problem Statement}
	
	In the practical scenario, it is difficult to know the precise location of the target and the perfect EUE CSI, so we assume that only a rough estimate of the target location and the imperfect CSI of BS-to-EUE link are available at the BS.
	In this case, the target location is roughly known among the angular uncertainty interval ${\bf{\Theta }} = \left[ {{\theta _0}  -  \Delta \theta ,{\theta _0}  +  \Delta \theta } \right]$.
	Besides, according to \cite{wang2021covert}, since there is always some inadvertent signal leakage from radiometer, the BS can estimate the imperfect CSI of BS-to-EUE link, which can be modeled as ${{\bf{h}}_{\rm{e}}} = {{\bf{\hat h}}_{\rm{e}}} + \Delta {{\bf{h}}_{\rm{e}}}$, where ${{\bf{\hat h}}_{\rm{e}}}$ is the channel estimate vector and $\Delta {{\bf{h}}_{\rm{e}}} \sim {\cal CN}( {0,\sigma _\upsilon ^2{\bf I}_{M_t}})$ is the estimation error vector.

	To realize a secure DFRC system by implementing the proposed I2S-aided HBF architecture, we propose to jointly design the I2S-aided HBF and radar receiver to maximize the minimum achievable LUE rate while meeting the requirements of EUE rate in the worst case, radar SINR, power and hardware constraints.
	Therefore, the HBF design problem is formulated as
	\begin{subequations}\label{eq:org}
		\begin{align}
		&\mathop {\max }\limits_{{\mathbf{A}},{{\mathbf{D}}_{\rm{C}}},{{\mathbf{d}}_{\rm{I}}},{\mathbf{w}}} \quad \mathop {\min }\limits_{u} \;{\rm{Rat}}{{\rm{e}}_u}\label{eq:org_a}\\
		&\;\;\quad\quad {\rm{s}}.{\rm{t}}.\quad \left\| {{\mathbf{A}}{{\mathbf{D}}_{\rm{C}}}} \right\|_F^2 + \left\| {{\mathbf{A}}{{\mathbf{d}}_{\rm{I}}}} \right\|_F^2 = \mathcal{P}\label{eq:org_b}\\
		& \qquad \qquad\quad \left| {{\mathbf{A}}[i,j]} \right| = 1,\forall i,j \label{eq:org_c}\\
		& \qquad \qquad \quad{\rm{ma}}{{\rm{x}}_{\Delta {{\bf{h}}_{\rm{e}}}}}{\rm{Rat}}{{\rm{e}}_{{\rm{e}},u}} \le {\xi _u},\forall u \label{eq:org_d}\\
		&\qquad \qquad \;\quad{\rm{SIN}}{{\rm{R}}_{\rm{r}}} \geqslant {\gamma _{\rm{r}}}\label{eq:org_e}
		\end{align}
	\end{subequations}%
	where $\mathcal{P}$ is transmit power budget, ${\xi _u}$ is EUE rate requirement in the worst case, and ${\gamma _{\rm{r}}}$ is radar SINR requirement.
	Besides, worst-case SR \cite{hota2022secure,al2023ergodic,sun2022ris} can be given by
	$
	\widetilde {{\rm{SR}}}= \textstyle \sum\nolimits_{u = 1}^U {{{[ {{\rm{Rat}}{{\rm{e}}_u} - \mathop {\max }_{\Delta {{\bf{h}}_{\rm{e}}}} {\rm{Rat}}{{\rm{e}}_{{\rm{e}},u}}} ]}^ + }} 
	$.

	\section{Hybrid Beamforming Algorithm}
	In this section, we reformulate the original problem into a more tractable form, and propose an AltOpt based algorithm to design receive filter and I2S-aided HBF.

	\subsection{Problem Reformulation}
	
	To address the uncertainty in the target angle, we divide the potential location set $\bf{\Theta }$ into $K$ discrete points uniformly, and then the radar SINR for $k$-th possible target angle is given by
	\begin{equation}\label{eq:sinr}
	\!\!{\rm{SIN}}{{\rm{R}}_{{\rm{r}},k}} = \frac{{\| {{\varsigma _0}{{\bf{w}}^H}{\bf{\tilde A}}({\theta _k}){{\bf{A}}}{{\bf{D}}_{\rm{CI}}}} \|_F^2}}{{\sum\nolimits_{i = 1}^I {\| {{\varsigma _i}{{\bf{w}}^H}{\bf{\tilde A}}({\theta _i}){{\bf{A}}}{\bf{D}}_{\rm{CI}}} \|_F^2}  + {{| {{{\bf{w}}^H}{\bf{z}}} |}^2}}}, \forall k
	\end{equation}%
	
	To tackle imperfect CSI of BS-to-EUE link, we construct a sample of i.i.d. EUE channel based on the set $\{{\bf h}_{\rm e}^n = {{\bf{\hat h}}_{\rm{e}}} + \Delta {{\bf{h}}_{\rm{e}}^n} | {{\bf{\hat h}}_{\rm{e}}}, \forall n \! \in \! \mathcal{N} \!\! = \!\! \{1,\dots,N\} \}$, where $\Delta {{\bf{h}}_{\rm{e}}^n}$ follows $\Delta {{\bf{h}}_{\rm{e}}^n} \sim \mathcal{CN}(0 , \sigma _\upsilon ^2{\bf I}_{M_t})$ and $N$ is the sample size.
	The EUE rate obtained over such a channel sample set can be defined as ${\rm{ {Rate}}}{_{{\rm{e}},u,n}}, \forall n \in \mathcal{N}$ \cite{hota2022secure,al2023ergodic,sun2022ris}.
	Accordingly, we can obtain
	\begin{equation}\label{eq:EUER2}
	{{\rm max} _{\Delta {{\bf{h}}_{\rm{e}}}}}\;{\rm{Rat}}{{\rm{e}}_{{\rm{e}},u}} = {{\rm max} _n}\;{\rm{Rat}}{{\rm{e}}_{{\rm{e}},u,n}},\forall u
	\end{equation}
	
	Based on the above reformulations, the original problem \eqref{eq:org} can be recast as
	\begin{subequations}
		\begin{align}
		&\mathop {\max }\limits_{  {{{\bf{A}}},{{\bf{D}}_{{\rm{C}}}},{{\bf{d}}_{{\rm{I}}}},{\bf{w}}} }\;\mathop {\min }\limits_u \;{\rm{Rat}}{{\rm{e}}_u}
		\label{10a}\\
		&\quad\;\;{\rm{s}}.{\rm{t}}.\quad\;\; \eqref{eq:org_b} \text{ and } \eqref{eq:org_c} \label{10b}\\
		&\qquad\qquad\;\;\;{{\rm max} _n} \; {\rm{Rate}}{_{{\rm{e}},u,n}} \le {\xi _u}, \forall u
		\label{10d}\\
		&\qquad\qquad\;\;\;{\rm{SIN}}{{\rm{R}}_{{\rm{r}},k}} \ge {\gamma _{\rm{r}}}, \forall k .
		\label{10e}
		\end{align}\label{10}%
	\end{subequations}%
	The variables $\{ {{{\bf{A}}},{{\bf{D}}_{{\rm{C}}}},{{\bf{d}}_{{\rm{I}}}},{\bf{w}}} \}$ are divided into receive filter $\mathbf{w}$ and I2S-aided HBF $\{\bf{A}, \bf{D}_{\rm{C}}, \bf{d}_{\rm{I}}\}$, which are designed in an AltOpt manner as detailed in the following subsections.

	\subsection{Receive Filter Optimization}
	
	With given I2S-aided HBF $\{\bf{A}, \bf{D}_{\rm{C}}, \bf{d}_{\rm{I}}\}$, the sub-problem for optimizing the receive filter $\bf{w}$ is given by
	\begin{equation}
	{\rm{Find}}\quad{\bf{w}}\quad {\rm{s.t.}}\;\; {{\rm{SINR}}_{{\rm r},k}}\ge \gamma _{\rm{r}}, \forall k
	\label{11}
	\end{equation}%
	To achieve better radar performance as possible, we formulate the problem \eqref{11}
	into a problem of maximizing radar SINR:
	\begin{equation}
	\mathop {\max }\limits_{\bf{w}} \;\mathop {\min } \limits_k \;\frac{{\| {{\varsigma _0}{{\bf{w}}^H}{\bf{\tilde A}}({\theta _k}){\bf{Y}}} \|_F^2}}{{\sum\nolimits_{i = 1}^I {\| {{\varsigma _i}{{\bf{w}}^H}{\bf{\tilde A}}({\theta _i}){\bf{Y}}} \|_F^2}  + {{| {{{\bf{w}}^H}{\bf{z}}} |}^2}}}
	\label{12}
	\end{equation}%
	According to \cite{Shen2018opt}, we can rewrite the max-min-ratio problem \eqref{12} into an equivalent form
	\begin{subequations}\label{W_1}
		\begin{align}
		&\!\!\!\!\!\!\!\mathop {\min }\limits_{{\bf{w}},{\bf{l}},c} \;  - c \label{W_1_a}\\	
		&\!\!\!\!\!\!\!\;\;{\rm{s.t}}. \;\;2{l_k}\sqrt {\| {{\varsigma _0}{{\bf{w}}^H}{\bf{\tilde A}}({\theta _k}){\bf{Y}}} \|_F^2}\nonumber\\
		&\!\!\!\!\!\!\!\qquad\;- l_k^2\!\left( {\textstyle\sum\nolimits_{i = 1}^I \!{\| {{\varsigma _i}{{\bf{w}}^H}{\bf{\tilde A}}({\theta _i}){\bf{Y}}} \|_F^2} \! +\! {{\left| {{{\bf{w}}^H}{\bf{z}}} \right|}^2}} \right) \!\!\ge c,\forall k \label{W_1_b}
		\end{align}
	\end{subequations}
	where ${\bf{l}} = [ {{l_1}, \ldots ,{l_K}} ]$ is a collection\footnote{  {$t$ is the iteration number, $( \cdot )^t$ denote the last point of $( \cdot )^{t+1}$.}} of variable $l_k$, $l_k^{t + 1}\!\! =\!\! \left(|{\varsigma _0}|\sqrt {\|{{({{\bf{w}}^t})}^H}{\bf{\tilde A}}({\theta _k}){\bf{Y}}\|_F^2} \right )/\left(\|{({{\bf{w}}^t})^H}{\bf{L}}\|_F^2 + {({{\bf{w}}^t})^H}{{\bf{w}}^t}\sigma _r^2\right)$, and ${\bf{L}} = [{\varsigma _1}{\bf{A}}({\theta _1}){\bf{Y}}, \ldots ,{\varsigma _I}{\bf{A}}({\theta _I}){\bf{Y}}]$.
	
	Note that the constraint \eqref{W_1_b} is non-convex, which is hard to tackle.
	To handle this problem, we adopt the Majorization-Minimization (MM) method and give the following lemma.
	\begin{lemma}\label{lemma1}
		Suppose $f({\bf{x}}) = {{\bf{x}}^H}{\bf{Qx}}$ with $\bf Q$ being Hermitian matrix, thus $f(\bf x)$ can be upper bounded as
		\begin{equation}
		f({\bf{x}}) = {{\bf{x}}^H}{\bf{Qx}} \le 2\Re \{ {({{\bf{x}}^t})^H}{\bf{Qx}}\}  - {({{\bf{x}}^t})^H}{\bf{Q}}{{\bf{x}}^t}
		\end{equation}
	\end{lemma}
	
	Based on \textit{{Lemma} \ref{lemma1}}, the problem \eqref{W_1} can be converted into
	\begin{equation}\label{W_2}
	\begin{aligned}
	&\mathop {\min }\limits_{{\bf{w}},{\bf{l}},c} \;  - c \\
	&\;\; {\rm s.t.} \; l_k^2( {\| {{{\bf{w}}^H}{\bf{L}}} \|_F^2 + {{\bf{w}}^H}{\bf{w}}\sigma _r^2} ) \\
	&\qquad- 2{l_k}| {{\varsigma _0}} |\sqrt { - \varkappa_k + 2\Re \{  {{\bf{\Lambda}}_k{\bf w}}\} } + c  \le  0 , \forall k\nonumber
	\end{aligned}
	\end{equation}
	where ${\bf{\Lambda }}_k = {({{\bf{w}}^t})^H}{\bf{\tilde A}}({\theta _k}){\bf{Y}}{{\bf{Y}}^H}{{{\bf{\tilde A}}}^H}({\theta _k})$ and $\varkappa_k =  \| {{{({{\bf{w}}^t})}^H}{\bf{\tilde A}}({\theta _k}){\bf{Y}}} \|_F^2$.
	This problem is convex and can be solved by existing solvers.
	
	\subsection{I2S-aided HBF Optimization}
	
	With given receive filter $\bf{w}$, the sub-problem for I2S-aided HBF can be expressed as
	\begin{equation}
	\mathop {\max }\limits_{ \substack{ {{{\bf{A}}},{{\bf{D}}_{{\rm{C}}}},{{\bf{d}}_{{\rm{I}}}},{\eta}} }} \quad \eta \qquad{\rm{s}}.{\rm{t}}. \;\; {\rm{Rat}}{{\rm{e}}_u} \ge \eta, \forall u, \eqref{10b} - \eqref{10e}
	\label{13}
	\end{equation}
	where $\eta$ is a newly introduced auxiliary variable.
	Problem \eqref{13} is hard to tackle due to the complex log-fractional term and coupled relationship among I2S-aided HBF.
	To deal with this, we give the following simplifications.
	
	\textit{1) Problem Simplifications:} To handle the complex term ${\rm{Rate}}_u$, we adopt the WMMSE-Rate relationship.
	Specifically, let ${\hat s_u}\!\! =\!\! {\kappa _u}{y_u}$ be the estimate of $s_u$ with the equalizer $\kappa _u$ for the $u$-th LUE. 
	Define MSE functions for the $u$-th LUE as ${\epsilon _u}$ $\! =\!{\mathbb E}\{ |{s_u} \!-\!{\kappa _u}{y_u}{|^2}\} \! =\!  | {{\kappa _u}} |^2  ( \sum\nolimits_{v = 1}^U\! {{| {{\bf{h}}_u^H{\bf{A}}{{\bf{d}}_{{\rm{C}},v}}} |}^2} \!{ + {{| {{\bf{h}}_u^H{\bf{A}}{{\bf{d}}_{\rm{I}}}} |}^2}} \!+  \sigma _u^2  ) $ $\!-\! 2\Re \{ {\kappa _u}{\bf{h}}_u^H\!{\bf{A}}{{\bf{d}}_{{\rm{C}},u}}\}   \!+\!  1$.
	The optimal unconstrained equalizer $\kappa _u$ for $u$-th LUE achieving minimum MSE is then given by
	\begin{equation}\label{14}
	{\kappa _u} = \frac{{{\bf{d}}_{{\rm{C}},{{u}}}^H{{\bf{A}}^H}{{\bf{h}}_u}}}{{\| {{\bf{h}}_u^H{\bf{A}}{{\bf{D}}_{{\rm{CI}}}}}\|_F^2 + \sigma _u^2}},\forall u
	\end{equation}
	and the corresponding minimum MSEs are $\epsilon_u^ \star  = {\min _{{\kappa _u}}}{\epsilon_u} = {(1 \!+\! {\rm{SIN}}{{\rm{R}}_{{{u}}}})^{ - 1}}$.
	Introducing a weight ${\omega _u}\!\!>\!\!0$ associating the MSE for $u$-th LUE, we can establish WMMSE-Rate relationship ${\rm{Rat}}{{\rm{e}}_u} \!=\! {\max _{{\omega _u}}}{\log _2}{\omega _u} \!- {\omega _u}\epsilon_u^ \star \! +\! 1$, where the maximum of the right-hand problem is achieved when weight ${\omega _u}$ satisfies
	\begin{equation}\label{15}
	\omega _u^ \star  = 1 + {\rm{SIN}}{{\rm{R}}_u},\forall u
	\end{equation}
	
	Then, we focus on tackling the coupled relationship among I2S-aided HBF.
	Note that $\bf A$, ${\bf{D}}_{\rm{C}}$ and ${\bf{d}}_{\rm{I}}$ are always coupled as ${\bf{A}}{\bf{D}}_{\rm{C}}$ and ${\bf{A}}{\bf{d}}_{\rm{I}}$, which motivates us to introduce two auxiliary variables ${\bf{Y}}_{\rm C} \!=\!\! [{\bf{y}}_1, \!\dots\!, {\bf{y}}_U] \!=\! {\bf{A}}{\bf{D}}_{\rm C}$ and ${\bf{y}}_0 \!= \!{\bf{A}}{\bf{d}}_{\rm I}$ to decouple.
	
	Based on the above formulations, the problem \eqref{13} can be rewritten as
	\begin{subequations}
		\begin{align}
		&\mathop {\max }\limits_{{\scriptstyle{\bf{Y}},\eta ,{\bf{A}},{{\bf{D}}_{{\rm{CI}}}},}{\scriptstyle\{ {\kappa _u}\} ,\{ {\omega _u}\} }} \;\eta \label{16a}\\
		&\qquad \quad\;\;{\rm{s}}.{\rm{t}}.\;\; - {\omega _u}{\epsilon_u}({\bf{Y}},{\kappa _u}) + {\log _2}{\omega _u} + 1 \ge \eta ,\forall u\label{16b}\\
		&\qquad \qquad\quad\;\;\| {\bf{Y}} \|_F^2 = {\cal P}\label{16c}\\
		&\qquad \qquad\quad\;\;| {{\bf{A}}[i,j]} | = 1,\forall i,j\label{16d}\\
		&\qquad \qquad\quad\;\; \max_n\; { \frac{{{{| {({\bf{h}}_{\rm{e}}^n)^H{{\bf{y}}_u}} |}^2}}}{{{{| {({\bf{h}}_{\rm{e}}^n)^H{{\bf{y}}_{0}}} |}^2} + \sigma _{\rm{e}}^2}}}  \le 2^{\xi _u}-1,\forall u\label{16e}\\
		&\qquad \qquad\quad\;\;\frac{{{{| {{\varsigma _0}} |}^2}\| {{{\bf{w}}^H}{\bf{\tilde A}}({\theta _k}){\bf{Y}}} \|_F^2}}{{\| {{\bf{BY}}} \|_F^2 + {{\bf{w}}^H}{\bf{w}}\sigma _r^2}} \ge {\gamma _{{\rm{r}},k}}, \forall k\label{16f}\\
		&\qquad \qquad\quad\;\;{\bf{Y}} = {\bf{A}}{{\bf{D}}_{{\rm{CI}}}}\label{16g}%
		\end{align}
	\end{subequations}
	where ${\bf{Y}} \!=\! [{{\bf{y}}_0},{{\bf{Y}}_{\rm{C}}}]$ and ${\bf{B}} \!=\! {[{\varsigma _1}{{\bf{w}}^H}{\bf{\tilde A}}({\theta _1}); \ldots ;{\varsigma _I}{{\bf{w}}^H}{\bf{\tilde A}}({\theta _I})]}$.
	
	Then, to deal with \eqref{16g}, we penalize equality constraint into objective function and obtain the following augmented Lagrangian (AL) minimization problem.
	\begin{equation}
	\mathop {\min }\limits_{{\scriptstyle{\bf{Y}},\eta ,{\bf{A}},}{\scriptstyle{{\bf{D}}_{{\rm{CI}}}}}} \;{\cal L}\left( {{\bf{Y}},\eta ,{\bf{A}},{{\bf{D}}_{{\rm{CI}}}}} \right)\quad {\rm{s.t.}}\; \eqref{16b}- \eqref{16f}
	\end{equation}
	where ${\cal L}({\bf{Y}},\eta ,{\bf{A}},{{\bf{D}}_{{\rm{CI}}}}) =  - \eta  + {\textstyle{\frac{\rho}{2}  }}||{\bf{Y}} - {\bf{A}}{{\bf{D}}_{{\rm{CI}}}} + {\bf{Z}}||_F^2$ is the AL function of this problem with $\rho>0$ being penalty parameter and $\bf{Z}$ being dual variable.
	
	\textit{2) Alternating Optimization of AL problem:} According to the above reformulations, ${\bf Y}$, $\bf A$ and ${\bf D}_{\rm CI}$ are separate, which can be alternatively updated by the following steps:
	
	\textit{Step 1:} Given other variables, $\bf Y$ and $\eta$ are updated by solving
	\begin{equation}
	\begin{aligned}
	&\mathop {\min }\limits_{{\bf{Y}},\eta } \; - \eta  + \frac{\rho }{2}\left\| {{\bf{Y}} - {{\bf{A}}}{{\bf{D}}_{{\rm{CI}}}} + {\bf{Z}}} \right\|_F^2\\
	&\;\;{\rm{s}}.{\rm{t}}.\;\; \eqref{16b},\eqref{16c},\eqref{16e}, {\rm{and}}\;\eqref{16f}
	\end{aligned}
	\label{18}
	\vspace{-0.5em}
	\end{equation}
	The problem \eqref{18} is hard to tackle because of the non-convex constraints \eqref{16e} and \eqref{16f}.
	
	Based on \textit{{Lemma} \ref{lemma1}}, we can reformulate \eqref{16e} and \eqref{16f} as
	\begin{equation}\label{20}
	{\bf{y}}_u^H{\bf{G}}_n{{\bf{y}}_u} - \Re \left\{ {{(\bf{g}}_0^{u,n})^H{{\bf{y}}_0}} \right\} + {\Upsilon _{u,n}} \le 0,\forall u, n
	\end{equation}
	\begin{equation}\label{21}
	( {\textstyle\sum\limits_{i = 0}^U\! {{{\left| {{\bf{B}}{{\bf{y}}_i}} \right|}^2}} \!\! + \! {{\bf{w}}^H}{\bf{w}}\sigma _r^2} ) - \Re \{ {{\rm{Tr}}\{ {\textstyle\sum\limits_{i = 0}^U \!{{\bf{M}}_k{{\bf{y}}_i}} } \}} \} + \Psi_k \! \le \! 0 ,\forall k
	\end{equation}%
	where ${\bf{G}}_n \!\! = \! {\bf{h}}_{\rm{e}}^n({\bf{h}}_{\rm{e}}^n)^H$, 
	${\Upsilon _{u,n}} \!= \!({2^{{\xi _u}}} - 1)({({\bf{y}}_0^t)^H}{{\bf{G}}_n\bf{y}}_0^t - \sigma _{\rm{e}}^2)$, 
	${{\bf{g}}_0^{u,n}} \! = \! 2({2^{{\xi _u}}} \! -\! 1){({\bf{y}}_0^t)^H}{\bf{G}}_n$,  ${\bf{J}}_k = {{{\bf{\tilde A}}}^H}({\theta _k}){\bf{w}}{{\bf{w}}^H}{\bf{\tilde A}}({\theta _k})$, 
	$ \Psi_k  = (|{\varsigma _0}{|^2}/{\gamma _{{\rm{r}},k}}){\rm{Tr}}\{ {{{({{\bf{Y}}^t})}^H}{\bf{J}}_k{{\bf{Y}}^t}} \}$ and ${\bf{M}}_k = {({{\bf{Y}}^t})^H}{\bf{J}}_k$.
	
	Based on the above reformulations, we obtain
	\begin{equation}
	\begin{aligned}
	&\mathop {\min }\limits_{{\bf{Y}},\eta } \; - \eta  + \frac{\rho }{2}\left\| {{\bf{Y}} - {{\bf{A}}}{{\bf{D}}_{{\rm{CI}}}} + {\bf{Z}}} \right\|_F^2\\
	&\;\;{\rm{s}}.{\rm{t}}.\;\; \eqref{16b}, \eqref{16c}, \eqref{20}, {\rm and}\;\eqref{21}
	\end{aligned}
	\end{equation}
	which is a conventional convex QCQP problem that can be solved by many existing solvers.
	
	\textit{Step 2:} Given other variables, $\bf{A}$ is updated by solving
	\begin{equation}
	\begin{aligned}
	&\mathop {\min }\limits_{{{\bf{A}}}} \;\left\| {{\bf{Y}} - {\bf{A}}{{\bf{D}}_{{\rm{CI}}}} + {\bf{Z}}} \right\|_F^2\\
	&\;\;{\rm{ s}}.{\rm{t}}.\;\left| {{{\bf{A}}}[i,j]} \right| = 1,\forall i,j 
	\end{aligned}
	\label{23}
	\end{equation}
	Problem \eqref{23} is a constant modulus constrained quadratic problem, which can be efficiently solved by element-wise BCD.
	Here, we omit the details of the element-wise BCD method and refer the reader to \cite{cheng2022hybrid}.
	
	\textit{Step 3:} Given other variables, ${\bf D}_{\rm CI}$ is updated by solving
	\begin{equation}
	\mathop {\min }\limits_{{{\bf{D}}_{{\rm{CI}}}}} \;\left\| {{\bf{Y}} - {{\bf{A}}}}{{\bf{D}}_{{\rm{CI}}}} + {\bf{Z}} \right\|_F^2
	\end{equation}
	whose close-form solution can be derived as
	\begin{equation}
	{{\bf{D}}_{{\rm{CI}}}} = {\left( {\bf{A}}^H{{\bf{A}}} \right)^{ - 1}}{\bf{A}}^H\left( {{\bf{Y}} + {\bf{Z}}} \right)
	\label{25}
	\end{equation}
	
	\subsection{Summary}
	
	According to the above derivations, the proposed secure hybrid beamforming (SHE) algorithm for joint design I2S-aided HBF and receive filter at DFRC BS is summarized in Algorithm \ref{alg1}.
	{Note that the proposed algorithm has double loops: 1) the outer loop is to update the   {receive filter} and the HBF; 2) the inner loop is to update the specific parameters of HBF, namely, ABF, communication DBF, and I2S DBF.}
	
	Then, we briefly analyze the computation complexity of the proposed SHE algorithm.
	Specifically, updating $\bf w$ requires the complexity of ${\cal O}(M_t^3)$.
	Updating $\{{\bf A}, {\bf D}_{\rm C}, {\bf d}_{\rm I}\}$ requires the complexity of ${\cal O}(N_{\rm AltOpt}({M_t^3}{M_r^3} + {N_{\rm BCD}}{M_t}{N_{\rm RF}^2} + {N_{\rm RF}^3}))$, where $N_{\rm AltOpt}$ and $N_{\rm BCD}$ denote the number of iterations of inner AltOpt algorithm and BCD algorithm, respectively.
	Therefore, the overall complexity is given by ${\cal O}({N_{\rm SHE}}({N_{\rm AltOpt}}({M_t^3}{M_r^3} + {N_{\rm BCD}}{M_t}{N_{\rm RF}^2} + {N_{\rm RF}^3}) + M_r^3 ))$, where ${N_{\rm SHE}}$ denotes the iteration number of the outer loop.

	\begin{algorithm}[!t]
		\small
		\caption{The proposed Secure Hybrid bEamforming (SHE) Algorithm for DFRC BS}
		\label{alg1}
		\begin{algorithmic}[1]
			\STATE \textbf{Input:} System Parameters.
			\WHILE{No Convergence}
			\STATE Update ${\bf w}^{[l]}$ by solving \eqref{12}.
			\WHILE{No Convergence}
			\STATE Update ${\bf Y}^{t+1}$ and ${\eta}^{t+1}$ by solving \eqref{18}.
			\STATE Update ${\bf A}^{t+1}$ by solving \eqref{23}.
			\STATE Update ${\bf D}_{\rm C}^{t+1}$ and ${\bf d}_{\rm I}^{t+1}$ by \eqref{25}.
			\STATE Update ${\{ {\kappa _u}\}}^{t+1}$ and ${\{ {\omega _u}\}}^{t+1}$ by \eqref{14} and \eqref{15}.
			\STATE   {${\bf Z}^{t+1} = {\bf Z}^{t} + {\bf Y}^{t+1} - {\bf A}^{t+1}{\bf D}_{\rm CI}^{t+1}$}.
			\ENDWHILE
			\STATE ${\bf A}^{[l]}={\bf A}^{t+1}$, ${\bf D}_{\rm C}^{[l]}={\bf D}_{\rm C}^{t+1}$, and ${\bf d}_{\rm I}^{[l]}={\bf d}_{\rm I}^{t+1}$.
			\ENDWHILE
			\STATE \textbf{Output:} ${\bf A}^{[l]}$, ${\bf D}_{\rm C}^{[l]}$, ${\bf d}_{\rm I}^{[l]}$, and ${\bf w}^{[l]}$.
		\end{algorithmic}
	\end{algorithm}

	\section{Simulation Examples}
	
	In this section, some simulation experiments are provided to validate the effectiveness of the proposed SHE algorithm and the superiority of the proposed I2S-aided HBF architecture.
	
	Unless otherwise specified, we assume that the DFRC BS is equipped with $N_{\rm RF} = 8$ RF chains, $M_t = 32$ transmit antennas, and $M_r = 8$ receive antennas, serving $U=4$ LUEs.
	The transmit power budget is set as ${\cal P} = 1$.
	For communication, we assume that the required EUE rates for all the LUEs are the same, i.e., ${\xi_u} = {\xi}, \forall u$.
	The communication noise power is set as $\sigma _u^2 = -10{\rm dB}, \forall u$ and the eavesdropping noise power is set as $\sigma_{\rm e}^2 =
	-20{\rm dB}$.
	For radar, we consider one target and $I=3$ interference sources and assume the angles of target and interference sources are   {${{\theta _0}} = 0^{\circ}$} and $\{ {\theta _i}\} _{i = 1}^I = -45^{\circ}, 30^{\circ}, 60^{\circ}$, respectively.
	Besides, the radar cross-section of target and interference sources are set as   {${{\varsigma _0}} = 10{\rm dB}$ and ${{\varsigma _i}} = 15{\rm dB},\forall i$}, respectively.
	The radar noise power is set as $\sigma_r^2 = -20 {\rm dB}$.
	{The number of discrete points on the angular grid $[-90^{\circ},90^{\circ}]$ is $361$, and the number of discrete points among possible target location region ${\bf{\Theta }}$ is $K = \Delta \theta / 0.5^{\circ}$.
		The sample size of EUE channel is $N = 20$.}
	{We set the measure of EUE channel estimation inaccuracy $\sigma ^2_\upsilon = 0.01$ and the uncertainty of target location is $\Delta \theta = 5^{\circ}$.}
	
	For comparison, we include the following benchmarks:
	1) DFRC I2S-aided FD-BF:
	the secure DFRC with I2S-aided fully digital (FD) BF is included as the upper bound of the DFRC with HBF architecture.
	2) DFRC Conv. HBF: 
	the DFRC with conventional HBF is introduced to show the superiority of the proposed I2S-aided HBF.
	3) Comm. Only I2S-aided HBF:
	the BS with I2S-aided HBF only works for downlink communication.
	4) Comm. Only Conv. HBF:
	the BS with conventional HBF only works for downlink communication.

\begin{figure}[t]
		\centering
		\includegraphics[height=0.6\linewidth]{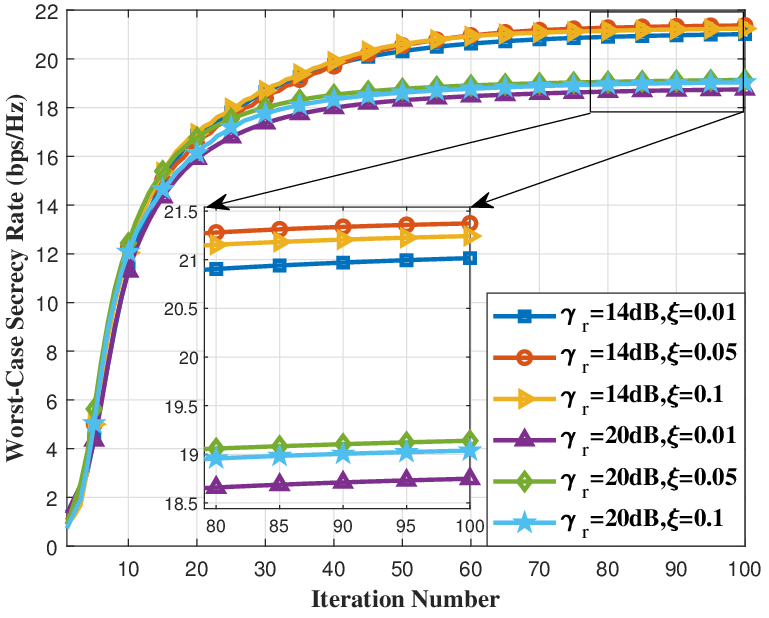} 
		\vspace{-0.5em}
		\caption{ The convergence performance: worst-case SR versus iteration number with different radar SINR threshold $\gamma_{\rm r}$ and EUE rate $\xi$.}
		\label{s-1}
		\vspace{-1em}
	\end{figure}

	
	Fig. \ref{s-1} plots the worst-case SR versus the iteration number with different radar SINR thresholds $\gamma_{\rm r}$ and EUE rate $\xi$.
	As can be observed, worst-case SRs are originally increasing and finally remain stationary for different $\gamma_{\rm r}$ and $\xi$.
	This validates the convergency of the proposed SHE algorithm.
	
\begin{figure}[t]
		\centering
		\includegraphics[height=0.6\linewidth]{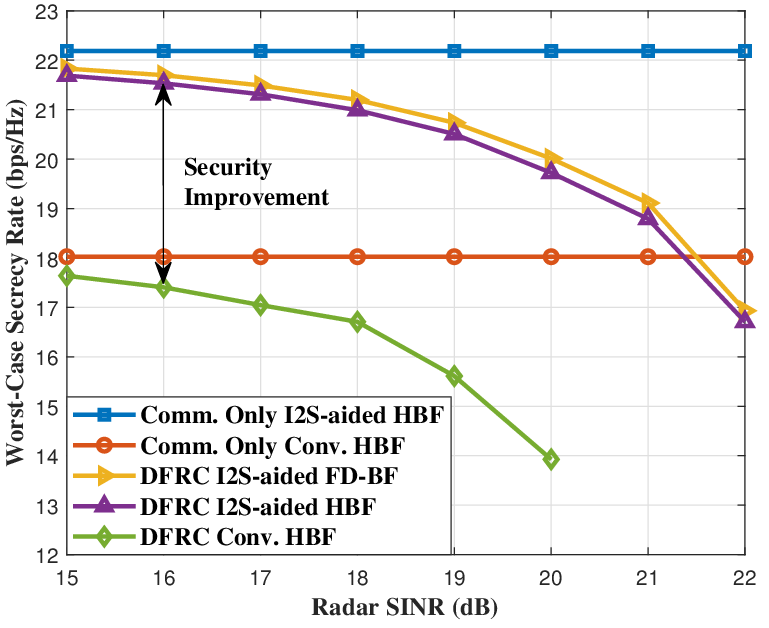}
		\vspace{-0.5em}
		\caption{The performance comparisons: worst-case SR versus radar SINR by applying different architectures.}
		\label{s-2}
		\vspace{-1em}
	\end{figure}

	
	In Fig.~\ref{s-2}, we compare the performance of the proposed architecture with other architectures versus the radar SINR $\gamma_{\rm r}$.
	It can be seen that all the curves corresponding to the DFRC BF experience a downward trend when increasing the radar SINR. 
	Noticeably, the curve of the worst-case SRs obtained by the proposed DFRC I2S-aided HBF approach to the curve of the DFRC I2S-aided FD-BF.
	Besides, we can observe that the proposed architecture achieves significantly higher SR than the DFRC Conv. HBF, and their gap becomes even larger with the radar SINR increasing.
	It should be also noted that worst-case SRs obtained by the proposed DFRC I2S-aided HBF are even higher than the Comm. Only Conv. HBF when the radar SINR is no more than $21{\rm dB}$.
	Therefore, it indicates the superiority of the proposed HBF architecture in PLS enhancement.

\begin{figure}[t]
		\centering
		\includegraphics[height=0.6\linewidth]{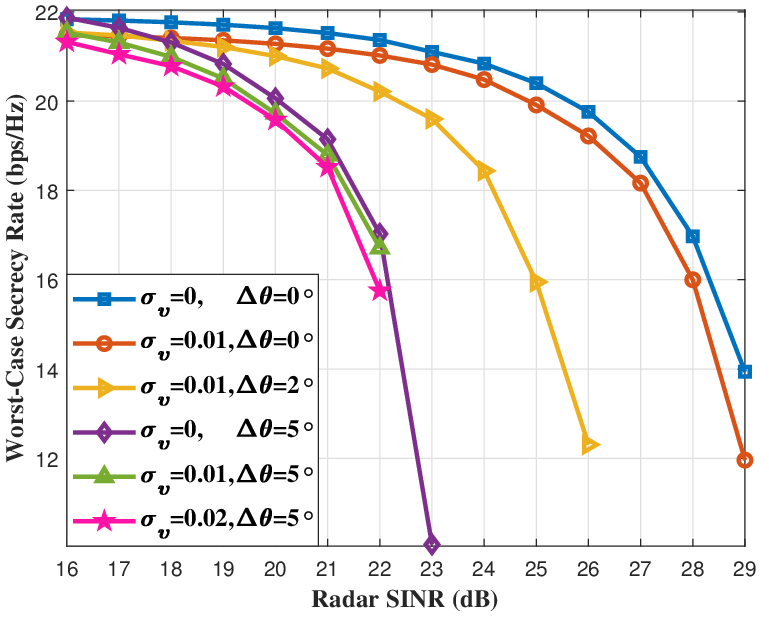}
		\vspace{-0.5em}
		\caption{Worst-case SR versus radar SINR with different target location uncertainty and imperfectly known EUE CSI.}
		\label{s-3}
		\vspace{-1em}
	\end{figure}

	In Fig.~\ref{s-3}, we plot the worst-case SRs versus radar SINR with different EUE channel estimation inaccuracy and target location uncertainty.
	It is clear that the worst-case SRs with a perfectly known radar target location and EUE CSI are the highest among all the conditions.
	With the same $\Delta \theta$, the worst-case SRs with higher $\sigma _\upsilon$ is lower.
	With the same $\sigma _\upsilon$, the worst-case SRs with larger $\Delta \theta$ decrease more dramatically.
	
	\section{Conclusion}
	
	In this letter, we propose an I2S-aided HBF architecture for the DFRC system to enhance PLS, where the radar target location and eavesdropper CSI are imperfectly known by the BS.
	In the proposed HBF architecture, we modify the DBF to introduce an extra I2S symbol, which can jam the eavesdropper and guarantee radar detection.
	The PLS enhancement problem is formulated by maximizing the minimum legitimate user communication rate with the constraints of eavesdropping rate, radar SINR, power and hardware.
	To tackle this problem, we devise a SHE algorithm based on the AltOpt method.
	Simulation results demonstrate that the proposed HBF algorithm effectively guarantees satisfactory performance of secure communication and radar target detection. 
	Moreover, the proposed HBF architecture outperforms traditional HBF in enhancing PLS.
	Future work can be conducted on a unified method for designing HBF with various imperfect CSI models.

	\footnotesize
	\balance
	
	\bibliographystyle{IEEEtran}
	\bibliography{IEEEabrv,Xu_WCL2023-2198}
	
\end{document}